\documentclass[aps,prl,amsmath,amssymb,twocolumn,nobibnotes,nofootinbib, showpacs, showkeys, nofootinbib]{revtex4}
\usepackage{pslatex,graphicx,dcolumn,bm,natbib}
\usepackage{amsmath,ulem,color}
\usepackage{newtxtext}
\usepackage[varvw]{newtxmath}
\usepackage{multirow}
\usepackage{subfigure}
\usepackage{color}{\tiny }
\usepackage[colorlinks,linkcolor=blue,anchorcolor=blue,citecolor=blue]{hyperref}
\newcommand{{\eee}}{\mbox{\Large\it e}}
\newcommand{{\HH}}{{\cal H}}

\setcitestyle{super}
\date{\today}

\begin{document}
\title{Disorder-induced vibrational anomalies from crystalline to amorphous solids}


\author{Ling Zhang$^{1,2}$, Yinqiao Wang$^2$, Yangrui Chen$^2$, Jin Shang$^2$, Aile Sun$^2$, Xulai Sun$^2$, Shuchang Yu$^2$, Jie Zheng$^2$, Yujie Wang$^2$, Walter Schirmacher$^{3}$, and Jie Zhang$^{2,4,\ast}$}
\affiliation{$^1$School of Automation, Central South University, Changsha 410083, China.}
\affiliation{$^2$School of Physics and Astronomy, Shanghai Jiao Tong University, Shanghai 200240, China}
\affiliation{$^3$Institut f$\ddot{u}$r Physik, Universit$\ddot{a}$t Mainz, Staudinger Weg 7, D-55099 Mainz, Germany}
\affiliation{$^4$Institute of Natural Sciences, Shanghai Jiao Tong University, Shanghai 200240, China}

\begin{abstract}
The origin of boson peak -- an excess of density of states over Debye's model in glassy solids -- is still under intense debate, among which some theories and experiments suggest that boson peak is related to van-Hove singularity. Here we show that boson peak and van-Hove singularity are well separated identities, by measuring the vibrational density of states of a two-dimensional granular system, where packings are tuned gradually from a crystalline, to polycrystals, and to an amorphous material. We observe a coexistence of well separated boson peak and van-Hove singularities in polycrystals, in which the van-Hove singularities gradually shift to higher frequency values while broadening their shapes and eventually disappear completely when the structural disorder $\eta$ becomes sufficiently high. By analyzing firstly the strongly disordered system ($\eta=1$) and the disordered granular crystals ($\eta=0$), and then systems of intermediate disorder with $\eta$ in between, we find that boson peak is associated with spatially uncorrelated random flucutations of shear modulus $\delta G/\langle G \rangle$ whereas the smearing of van-Hove singularities is associated with spatially correlated fluctuations of shear modulus $\delta G/\langle G \rangle$.
\end{abstract}


\keywords{Boson peak, van Hove singularities, granular materials}
-\pacs{83.80.Fg, 45.70.-n, 81.05.Rm, 61.43.-j}

\maketitle

Glassy materials show anomalies with respect to Debye's predictions, for example, a prominent peak in the specific heat relative to $T^3$ \cite{zellerpohl,1988-Y.Leggett-CommCondMatPhys,1997-AP.Sokolov-PRL}, a plateau in the temperature variation of the thermal conductivity \cite{1986-JJ.Freeman-AC.Anderson-PRB, 2006-W.Schirmacher-EL}, and a remarkable peak in the reduced density of states $D(\omega)/\omega^{d-1}$, normalized with respect to the Debye law, with $d$ being the dimension \cite{Buchenau-PRB-1986, 1986-AP.Sokolov-SolidStateCom,1998-W.Schirmacher-PRL,2008-H.Tanaka-NatureMat}. This latter peak is called "boson peak" (BP) and is closely related to the mentioned other thermal anomalies.
BP is associated with a strong increase of the sound attenuation \cite{2009-G.Monaco-PNAS,2013-W.Schirmacher-ScientificRep} as well as a minimum in the sound velocity \cite{2009-G.Monaco-PNAS,2013-W.Schirmacher-ScientificRep}. Near BP frequency, sound waves reach the Ioffe-Regel limit \cite{2008-H.Tanaka-NatureMat}.

To understand the origin of the BP-related vibrational anomalies has been taken as a challenge by many workers ever since their discoveries \cite{zellerpohl,Jackle1981,Buchenau-PRL-1984, Buchenau-PRB-1986,1986-AP.Sokolov-SolidStateCom,1991-Laird-PRL,1998-W.Schirmacher-PRL,Baggioli2019universal,Baggioli2019vibrational}.
In most efforts, the structural disorder has been considered crucial. In theories, the influence of disorder on vibrational spectrum has been modelled by spatially fluctuating potentials \cite{karpov83,Buchenau91a,gurevich03a}, force constants 
\cite{1998-W.Schirmacher-PRL,martin-mayor00,2001-SN.Taraskin-PRL} and elastic constants \cite{2006-W.Schirmacher-EL,2007-W.Schirmacher-PRL,2013-W.Schirmacher-ScientificRep}. In particular, excellent agreement between theory and both experiments and simulations has been achieved using effective-medium theories \cite{2013-W.Schirmacher-ScientificRep,2013-W.Schirmacher-PSSB,schi14}.

However, there is still controversial and opposite opinion that BP would be a washed-out version of the lowest van-Hove singularity (VHS) of the transverse phonon branch
\cite{2001-SN.Taraskin-PRL,2011-Chumakov-PRL,2014-Chumakov-PRL,zorn11}. The argument in favor of this interpretation is that in lattice models of force-constant disorder the VHS peak smoothly transforms into the boson peak with increasing disorder \cite{2001-SN.Taraskin-PRL,zorn11} and that the BP in some glasses appears in the same frequency regime as the VHS peak of the corresponding crystal, if re-scaled with the Debye frequency. A rather strong argument against this interpretation is that near and beyond the BP frequency the Ioffe-regel limit shall be reached and a dispersion of the transverse branch can not be defined \cite{2008-H.Tanaka-NatureMat,Wang2018-disentangling}.

To shed new light on this important issue, it is desirable to look into more pieces of experimental and theoretical evidence. Here we present such a piece of evidence by means of a two-dimensional (2D) experimental system of granular materials in which the structural order can be tuned from the limit of crystal to glass. We find that BP and VHS are clearly separated.
Specifically, we observe the coexistence of the BP and VHS in polycrystals of moderate degrees of disorder. Strikingly, we also observe that when the degree of disorder increases, the VHS shift to higher frequencies. These observations strongly suggest that in general there is no generic connection between the BP and the lowest VHS, in contrast to the propositions in literature\cite{2001-SN.Taraskin-PRL,2011-Chumakov-PRL,2014-Chumakov-PRL}. In addition, we investigate the relationship between the BP and the relative fluctuations of shear modulus $\delta G/\langle G \rangle$ and the structural disorder $\eta$. We find that boson peak is associated with spatially uncorrelated random flucutations of shear modulus $\delta G/\langle G \rangle$ whereas the smearing of van-Hove singularities is associated with spatially correlated fluctuations of shear modulus $\delta G/\langle G \rangle$ that are reminiscent of crystalline solid and eventually vanish when $\eta$ is sufficiently high.

\begin{figure}
\centerline{\includegraphics[trim=0cm 0cm 0cm 0cm, width=1.0\linewidth]{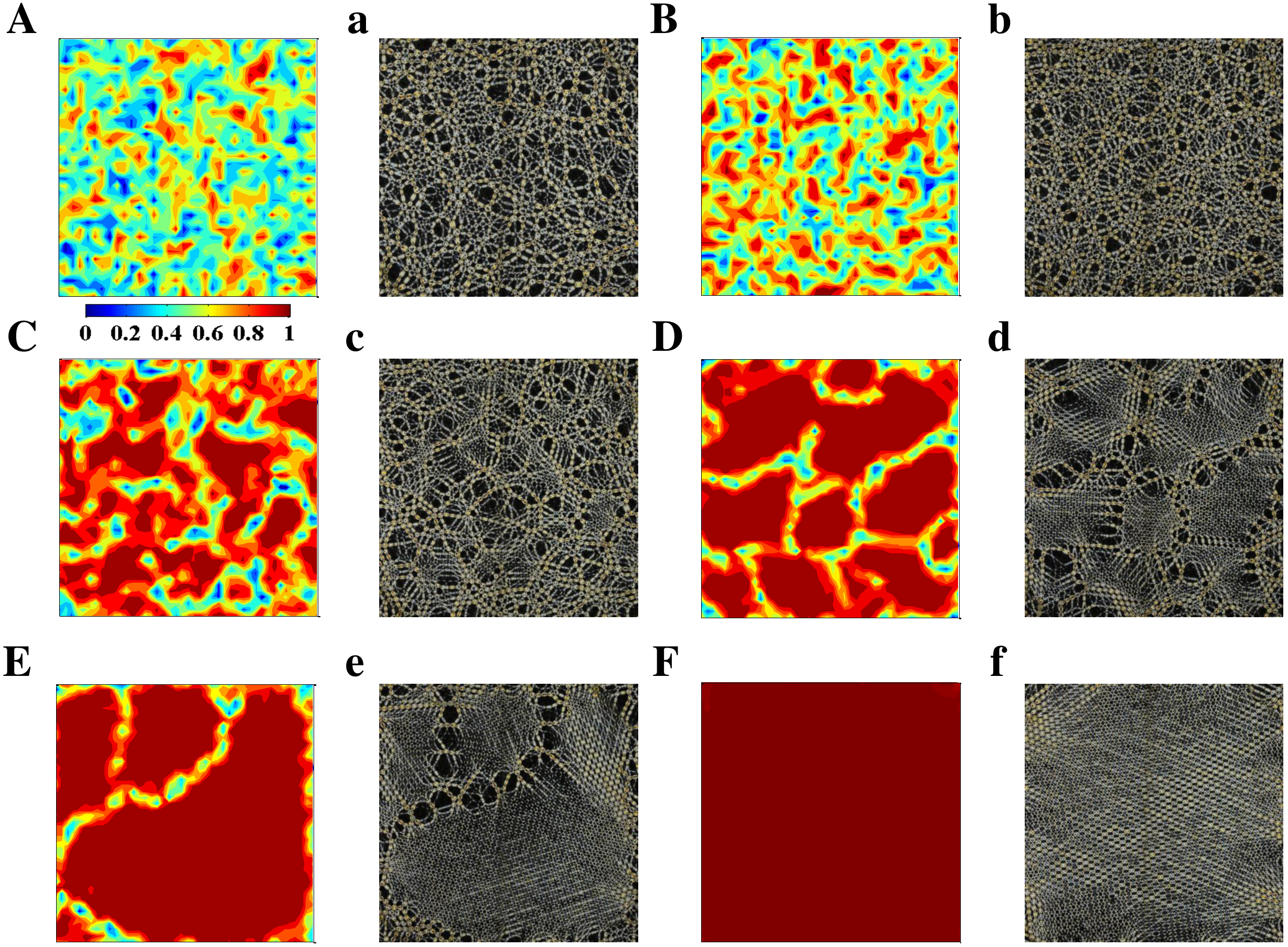}}
\caption{\label{packings} Images of bond orientation parameter $\psi^i_6$ (A-F) and force chains (a-f) of packings of various disorders. The same letters in upper and lower cases denote the same packing. The disorder is quantified using a number ratio between small and large disks: 1:1($\eta=1$, A,a), 6.25:1($\eta=0.74$, B,b), 33.1:1($\eta=0.41$, C,c), 178.92:1($\eta=0.2$, D,d), 1:0 in a poly-crystal ($\eta=0.13$, E,e), and 1:0 in a disordered crystal ($\eta=0$, F,f), respectively. Each packing has $\sim 2000$ disks and the pressure $p=26.5\pm0.2$ N/m.}
\end{figure}

In this experiment, we study the vibrational properties of 2D dense packings of photoelastic disks with various degrees of structural disorder, ranging from the strongly disordered system to granular crystals of weak force disorder, by systematically changing the number ratio of small disks (diameter 1.0 cm) to large disks (1.4cm). For a ratio of 1:1, we obtain the maximum random packing, whereas in the pure small-disk system the packing forms a crystal of triangular lattice. 
We use the bond-orientation parameter $\psi^i_6$ to determine the local hexagonal order, as depicted in panels $A-F$ in Fig.~\ref{packings}. The $\psi^i_6$ is defined as \cite{2011-K.Binder-W.Kob-GlassBook,2013-K.Chen-PRE,2010-H.Tanaka-NatrueMat}
$\psi^i_6=\left| \frac{1}{N_{nn}}\sum_k^{N_{nn}}e^{6j\theta_{ik}} \right|$,
where $N_{nn}$ is the number of nearest neighbors, $\theta_{ik}$ is the bond angle between disk $i$ and disk $k$, and $j\equiv\sqrt{-1}$. The average $\psi^i_6$ varies between the values of $\langle\psi^i_6\rangle=0.552$ (for a ratio 1:1) and $\langle\psi^i_6\rangle=1$ (for the crystal). In samples of intermediate composition of small and large disks, $\langle\psi^i_6\rangle$ varies continuously. We define a disorder parameter $\eta=(1-\langle\psi^i_6\rangle)/0.448$, such that $\eta=0$ for crystal and $\eta=1$ for the maximum random packing.

For a given $\eta$, we repeat experiment 10 times starting from an independently prepared initial packing. Images of the spatial distribution of $\psi^i_6$ and force-chains of 6 different $\eta$ are shown in Fig.~\ref{packings}, including the polycrystals in panels (C,c), (D,d), and (E,e) and a crystal packing with a weak disorder of contact forces in panels (F,f). To create an initial packing, we use a ``biax" \cite{2011-JZ-Nature,2017-LZhang-NatCommun,Wang2018-disentangling}, where 4 walls of a square domain move symmetrically while its center is fixed. This biax is filled with photoelastic disks, from $\sim$ 1500 to $\sim$ 2200 depending on $\eta$. Starting from an unjammed state, we compress the system isotropically till it approaches jamming, e.g. its volume fraction $\phi\sim0.84$ if $\eta=1$, and $\phi\sim0.90$ if $\eta=0$. During the compression, we constantly apply agitations to eliminate residual force chains to produce a random close packing (R.C.P.) of bidisperse particles of $\eta=1$ (size ratio 1:1.4 and a number ratio 1:1), resembling the corresponding R.C.P. of frictionless particles\cite{2006-Ohern-PRE,2005-LE.Silbert-PRL}. We prepare initial packing of other $\eta$ in a similar way. Starting from an initial packing, we compress the system into a set of highly jammed packings in a series of steps, where the strain increment in each step is $4.26\times10^{-4}$. At each step, images of particle configurations and force chains are taken for further analyses. After a precise measurement of contact forces between disks \cite{2011-JZ-Nature,2017-LZhang-NatCommun,Wang2018-disentangling}, we construct Hessian matrices to analyze the normal-mode vibrations of the set of packing\cite{2017-LZhang-NatCommun, Wang2018-disentangling}. Using Hessian matrices, we then determine the DOS and reduced DOS of these systems \cite{2017-LZhang-NatCommun, Wang2018-disentangling}.

\begin{figure}
	\centerline{\includegraphics[trim=0cm 0cm 0cm 0cm, width=1.0\linewidth]{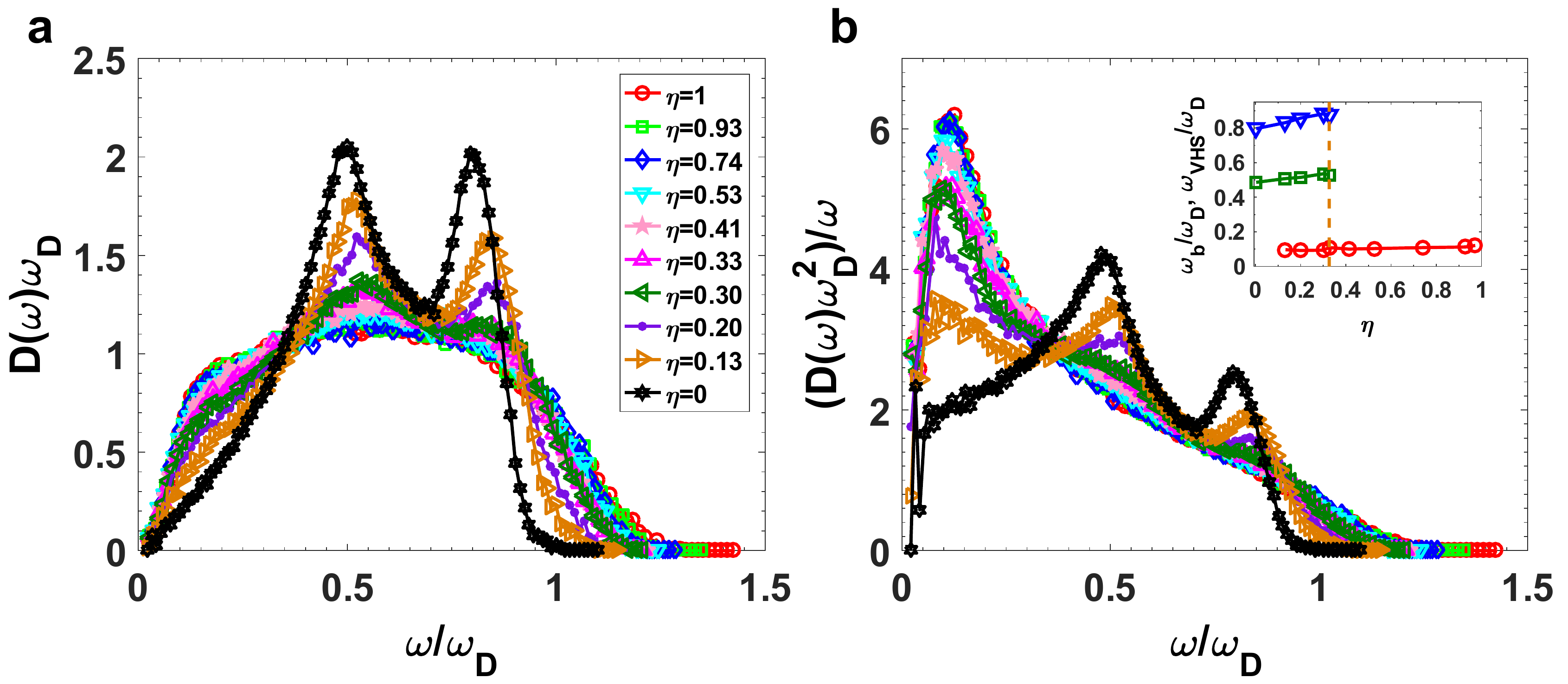}}
	\caption{\label{DOS} The DOSs \textbf{(a)} and reduced DOSs \textbf{(b)} of packings of different disorder parameters $\eta$. Here each curve is ensemble averaged over 10 independent packings of the same $\eta$. Moreover, the frequency $\omega$ is normalized by the Debye frequency $\omega_D$. Inset of \textbf{(b)}: The positions of the boson peak $\omega_b/\omega_D$ (red $\circ$), the first and second van Hove singularities (VHS) -- $\omega_{vH1}/\omega_D$ (green $\square$) and $\omega_{vH2}\omega_D$ (blue $\bigtriangledown$) v.s. $\eta$. The dashed line denotes the minimum $\eta$ beyond which the two VHS disappear. Here $p=26.5\pm0.2$ N/m.}
\end{figure}

The density of states (DOS) and reduced DOS of packings of different $\eta$ are shown in Fig.~\ref{DOS}, where we normalize the frequency $\omega$ using Debye's frequency $\omega_D$ -- a natural frequency unit, and we let $\int D(\frac{\omega}{\omega_D}) d(\frac{\omega}{\omega_D})=1$. 
When $0< \eta \leq 0.33$, on each curve, a shoulder in DOS or a corresponding BP appears in reduced DOS in the low frequency regime, and, meanwhile, two VHSs appear in the high frequency regime, as seen in Fig.~\ref{DOS} (a-b). 
When $\eta>0.33$, two VHSs barely exist with two small bumps reminisent of VHSs, as shown in Fig.~\ref{DOS} (a-b). The evolution  in Fig.~\ref{DOS} shows a clear level repulsion due to the gradual loss of symmetry from crystal to more and more disordered systems, which pushes the states of two VHSs into the low frequency regime -- leading to the formation of BP -- and simultaneously into the high frequency regime -- causing Anderson localization \cite{1958-PW.Anderson-PhyscalReview, 1998-W.Schirmacher-PRL,2008-HefeiHu-NatPhys, 2009-BJ.Huang-TM.Wu-PRE, 2021-LZhang-Level}, in excellent agreement with the theoretical propositions \cite{1998-W.Schirmacher-PRL,2012-W.Schirmacher-J.PhysCondensMatter, 2012-W.Schirmacher-EPL}. 


In addition, a striking characteristic is that two VHSs shift gradually to higher frequencies as $\eta$ increases, as shown in Fig.~\ref{DOS} (b) and its inset. The above observations are in sharp contrast to Ref.\cite{2001-SN.Taraskin-PRL} where the lowest VHS gradually shifts to lower frequencies as the disorder of force constant increases, leading to the proposition\cite{2001-SN.Taraskin-PRL} that the BP is generically related to the lowest VHS. Moreover, the frequency of lowest VHS  and the the BP frequency are close with a ratio of $1.4\sim2$ for sufficiently high disorders of force constant~\cite{2001-SN.Taraskin-PRL}, whereas these two frequencies are rather distinct with a ratio of $\sim5$ as shown in Fig.~\ref{DOS} (b). The opposite trends in the change of VHS may be understood from the different nature of disorder between Ref.~\cite{2001-SN.Taraskin-PRL} and our system: Ref.~\cite{2001-SN.Taraskin-PRL} is a lattice model with the disorder of force constant, different from the structure disorder in our system. These two different types of disorder have long been believed to be equivalent \cite{2001-SN.Taraskin-PRL}, but have qualitative differences in causing the evolution of VHS, as confirmed in the above discovery.

In the simulations \cite{2015-HTong-NXu-SciRep, 2017-YHNie-FrontPhys}, the authors also observed the coexistence of BP and VHSs in a weakly disordered 2D system, by systematically increasing the structural disorder starting from a perfect crystal. However, there are two differences between their findings and our experimental results, as follows. First, in simulations, it is easy to isolate disorder type one at a time, by exploring force constant disorder on a perfec lattice\cite{2017-YHNie-FrontPhys, 2015-HTong-NXu-SciRep} and lattice disorder \cite{2017-YHNie-FrontPhys}, poly-dispersity of particles \cite{2015-HTong-NXu-SciRep}, the fluctuations of local coordination number, or vacancies \cite{2017-YHNie-FrontPhys} to sufficiently high degrees. In our experiment, it is inevitable to incorporate all different types of disorders considered in Refs\cite{2015-HTong-NXu-SciRep, 2017-YHNie-FrontPhys}. Moreover, the positions of the lowest VHS hardly change with disorder \cite{2015-HTong-NXu-SciRep, 2017-YHNie-FrontPhys}, while in our findings, the lowest VHS evidently shifts to higher frequencies, as seen in Fig.~\ref{DOS}, which further verifies the distinction between BP and VHS.

Apparently, here BPs are necessarily induced by disorder, including both the structural disorder and the force-constant disorder. To disentangle the two, we firstly investigate the role of the force-constant disorder in the strongly disordered system ($\eta=1$) and the disordered granlar crystals ($\eta=0$). Here, we adopt the relative fluctuations of shear modulus, $\delta G/\langle G \rangle$,  which is confirmed to be the leading contribution to BP formation \cite{2017-LZhang-NatCommun}.  

\begin{figure}
	\centerline{\includegraphics[trim=0cm 0cm 0cm 0cm, width=1.0\linewidth]{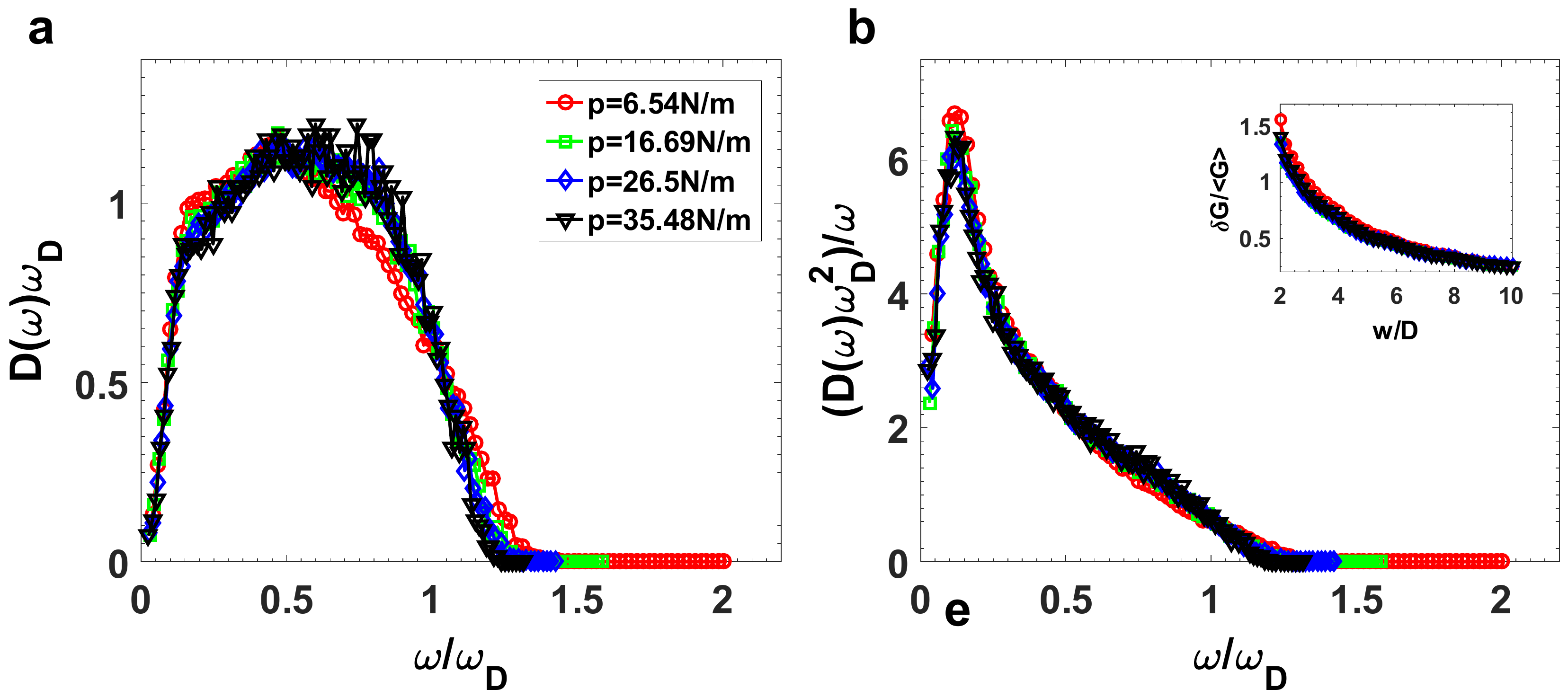}}
	\caption{\label{D(w)_eta1} The DOS {\bf(a)} and  the reduced DOS {\bf(b)} at different p collapse for the strongly disordered system($\eta=1$). Only data points from 4 different p are shown for clarity, results at other p are similar. Inset of {\bf(b)}:  The relative fluctuations of shear modulus, $\delta G/\langle G \rangle$ v.s. $w/D$ at different pressure. Results are independent
	of p.}
\end{figure}

As shown in Fig.~\ref{D(w)_eta1}(a-b),  the normalized DOSs and the reduced DOSs do not show significant changes in strongly disordered systems ($\eta=1$) when pressure $p$ changes over five times.  The measured $\delta G/\langle G \rangle$ at different pressure show little pressure dependence, as depicted in the inset of Fig.~\ref{D(w)_eta1}(b), in consistent with the results in the main panels of (a) and (b).

\begin{figure}
	\centerline{\includegraphics[trim=0cm 0cm 0cm 0cm, width=1.0\linewidth]{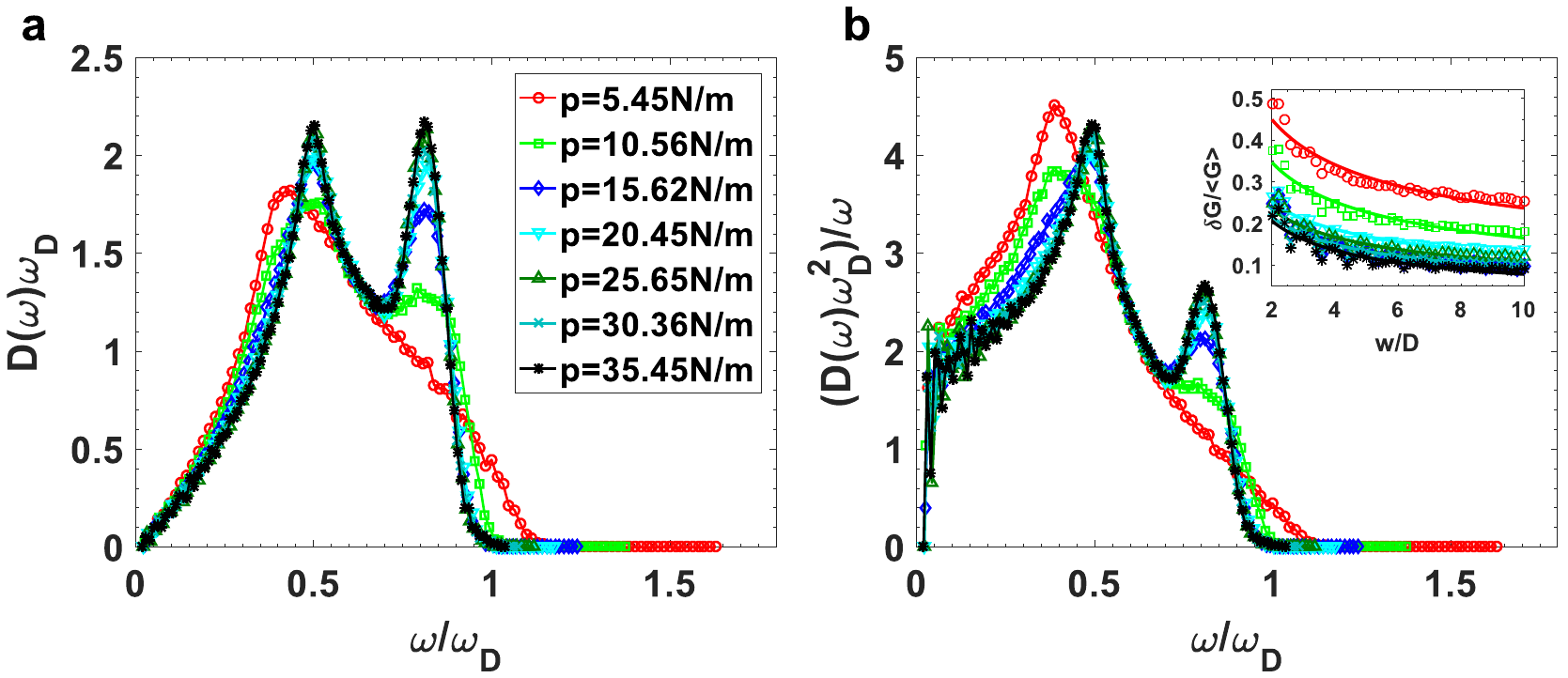}}
	\caption{\label{D(w)_Cryst}  The DOS {\bf(a)} and the reduced DOS {\bf(b)} rescaled by the Debye frequency $\omega_D$ at different pressure.  Each curve in {\bf(a,b)} is ensemble averaged over ten different runs at the same pressure. }
\end{figure}

Figure~\ref{D(w)_Cryst} (a) shows the normalized DOSs at different pressure $p$ in the disordered granular crystals, where the triangular-lattice structure is fixed but the force constant is disordered, due to the force network and the nonlinear interactions \cite{2008-BP.tighe-PRL}: when $p\leq5.45 N/m$, only the transverse VHS is left followed with a knee structure that gradually transforms to the longitudinal VHS when $p>5.45 N/m$. Correspondingly in Fig.~\ref{D(w)_Cryst}(b), only the transverse VHS is followed by a knee structure in the reduced DOSs, nearly invisible at $p=5.45 N/m$.  In contrast, the transverse VHS changes little after normalizing $\omega$ by $\omega_D$, which further verifies that BP and VHS are two different identities. Additionally, when $p>15.62 N/m$, both DOSs and reduced DOSs show an approximate collapse, as shown in Figure~\ref{D(w)_Cryst} (a-b).

To understand the results in Fig.~\ref{D(w)_Cryst}, we also measure $\delta G/\langle G \rangle$, as shown in the inset of Figure~\ref{D(w)_Cryst} (b). Apparently, there is a change in $\delta G/\langle G \rangle$ when $p\leq15.62 N/m$, in consistent with the evolution of the DOSs and reduced DOSs, especially the longitudinal VHS. When $p>15.62 N/m$, $\delta G/\langle G \rangle$ is not dependent on $p$ any more, consistent with the behavior of the DOSs and reduced DOSs. 

Comparing the insets of Fig.~\ref{D(w)_eta1}(b) and Fig.~\ref{D(w)_Cryst}(b), $\delta G/\langle G \rangle$ has a much larger value in Fig.~\ref{D(w)_eta1}(b) at the particle scale compared to that of Fig.~\ref{D(w)_Cryst}(b), and then starts to decay rapidly as $\delta G/\langle G \rangle\propto w^{-1.0\pm0.05}$ as shown in Fig.~\ref{D(w)_eta1}(b), with no long-range correlation in space \cite{2017-LZhang-NatCommun}, whereas $\delta G/\langle G \rangle\propto w^{-0.54\pm0.05}$ decays much slower in Fig.~\ref{D(w)_Cryst}(b) with spatial correlations -- reminiscent of crystalline structure, suggesting the qualitative difference in two cases. Hence, the Gaussian random fluctuations of shear-mudulus play the crucial role in the BP formation.

\begin{figure}
\centerline{\includegraphics[trim=0cm 0cm 0cm 0cm,width=1.0\linewidth]{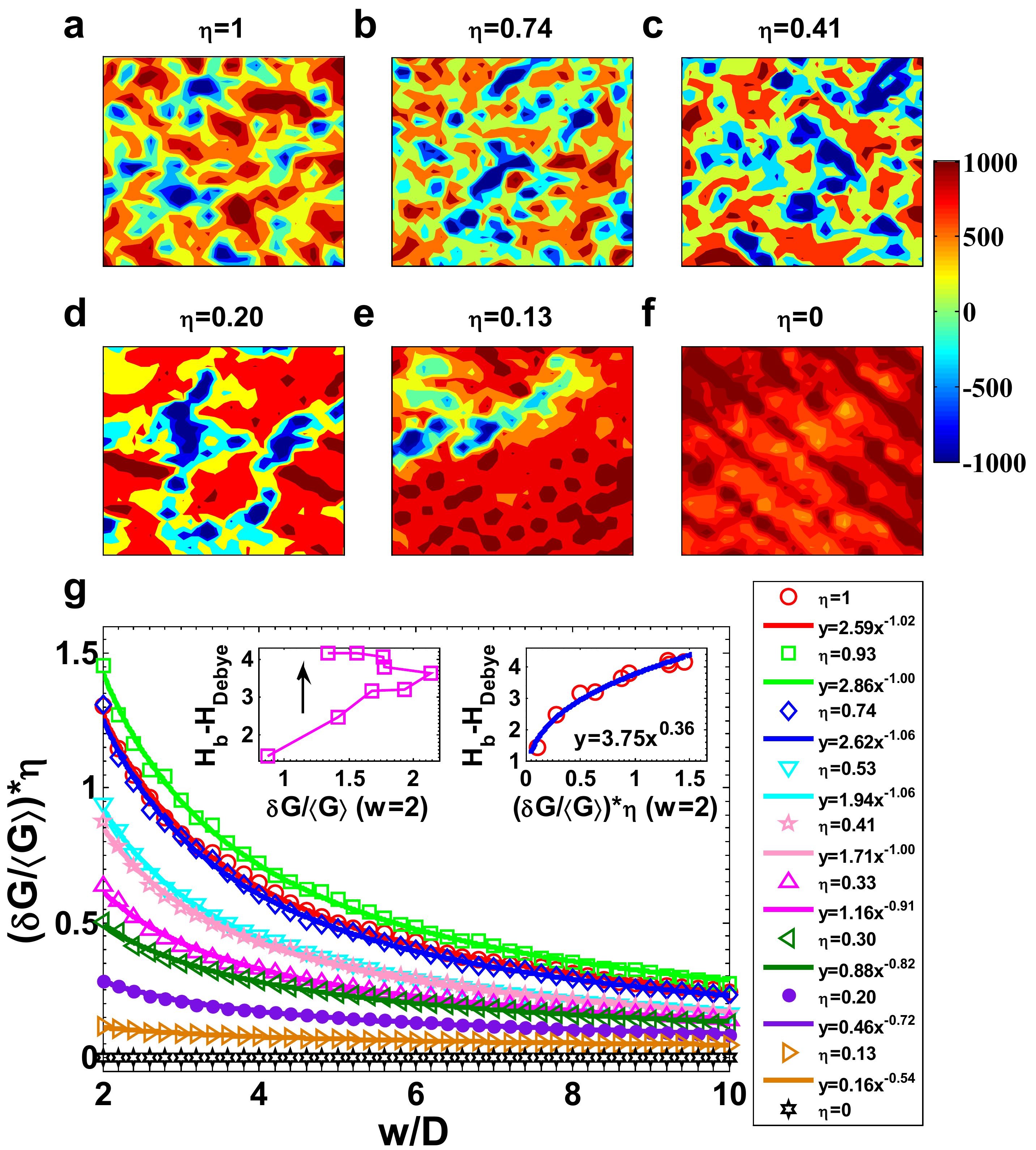}}
\caption{\label{deltaG_eta} {\bf (a-f)} The spatial distributions of local shear modulus of packings of different $\eta$. {\bf (g)} The average relative fluctuations of the shear modulus $\frac{\delta G}{\langle G \rangle}*\eta$ v.s. $w$ for different $\eta$. Solid lines are power law fit, as shown in the legends. Here $p=26.5N/m$, and the results of other p are similar. Inset: the height of the BP, $H_b$, relative to Debye level $H_{Debye}$ v.s. $\delta G/\langle G \rangle$ (left), where the arrow denotes the increment of $\eta$, and $\delta G/\langle G \rangle * \eta$ (right), where the blue solid line, $y=3.75x^{0.36}$, is a guide to eyes.}
\end{figure}

How about the role of structural disorder $\eta$? To better understand Fig.~\ref{DOS}, we firstly plot the spatial fluctuations of shear modulus $G$ of packings of different $\eta$, as shown in Fig.~\ref{deltaG_eta}(a-f). There are local regimes of negative shear modulus $G<0$ -- an indication of local softness -- in packings of all $\eta$ except in the disordered crystal where $\eta=0$, as shown in Fig.~\ref{deltaG_eta}(a-f). Figure~\ref{deltaG_eta}, together with Fig.~\ref{DOS}, suggests that the formation of the BP is closely related to local soft regimes in space, in good agreement with the early propositions \cite{2013-W.Schirmacher-ScientificRep, 2008-H.Tanaka-NatureMat, 1991-Laird-PRL, 2017-LZhang-NatCommun}. Recall that when $\eta>0.33$, the two VHS vanish, as shown in Fig.~\ref{DOS}. Coincidentally, the relative fluctuations of shear modulus $\delta G/\langle G \rangle$ also show a qualitatively change as a function of the coarse-graining size $w$ at $\eta=0.33$, as shown in Fig.~\ref{deltaG_eta}(g). When $\eta>0.33$, $\delta G/\langle G \rangle$ behave as a Gaussian random field with the power-law scaling exponent $\approx1.0$. A scaling exponent of 1.0 indicates that these relative fluctuations of $G$ are short-range correlated and their statistics obey the central limit theorem, which is clearly induced by the increase of structure disorder $\eta$. In addition, $\delta G/\langle G \rangle$ is the leading contribution, which is at least four times as large as the bulk modulus (data not shown), as reported in our early study\cite{2017-LZhang-NatCommun}, consistent with theories \cite{1998-W.Schirmacher-PRL,2008-W.Schirmacher-PSSB,2013-W.Schirmacher-PSSB,2013-W.Schirmacher-ScientificRep}. In contrast, when $\eta \leq 0.33$, the exponents systematically deviate from $1.0$ as $\eta$ decreases due to the spatial correlations of the fluctuations of $G$, as shown in the legend of Fig.~\ref{deltaG_eta}(g). Interestingly, we find an empirical relation that the relative height of the BP -- ($H_b-H_{Debye}$) as measured in Fig.~\ref{DOS}, increases monotonically as a power-law function of $\delta G/\langle G \rangle \eta$ with an scaling exponent of $0.36$, as shown in the inset (right) of Fig.~\ref{deltaG_eta}(g). Presently, we do not have a theoretical explanation for this exponent.



In conclusion, we study the vibrational properties of dense granular materials with a broad range of structural disorders and find that BP and VHS are two well separated features. We observe the coexistence between BP and VHS in packings of moderate degrees of disorder with $\eta \leq 0.33$ and the complete disappearance of two VHS when $\eta>0.33$. Around $\eta=0.33$, there is an associated qualitatively difference in the spatial flucutions of shear modulus. 

By analyzing fluctuations of shear modulus, $\delta G/\langle G \rangle$, at different pressure $p$ both in the strongly disordered system ($\eta=1$) and granular crystals ($\eta=0$), we find the Gaussian random spatial fluctuations of $\delta G/\langle G \rangle$  play a crucial role in the formation of BP \cite{2013-W.Schirmacher-ScientificRep,2013-W.Schirmacher-PSSB,2017-LZhang-NatCommun,Wang2018-disentangling}.

We further find that the formation of BP is closely related to local soft regimes of negative shear modulus, as found in packings of all $\eta$ except in the crystal. Moreover, we find that $\delta G/\langle G \rangle$ fluctuate randomly in space as Gaussian random variables when $\eta > 0.33$, below which its fluctuation has spatial correlation and two VHSs become visible. Hence, the formation of BP is accompanied with spatially uncorrelated random fluctuations of shear modulus and the evolution of VHS are accompanied with spatially correlated fluctuations of shear modulus. 

\begin{acknowledgments}
	This work is supported by the National Natural Science Foundation of China (NSFC) under (No.11774221 and No. 11974238). This work is also supported by the Innovation Program of Shanghai Municipal Education Commission under No 2021-01-07-00-02-E00138. LZ also thanks the support by NSFC under No.11904410. We also acknowledge the support from the student innovation center of Shanghai Jiao Tong University.
\end{acknowledgments}
$^{\ast}$(jiezhang2012@sjtu.edu.cn)


\begin{thebibliography}{45}
	\expandafter\ifx\csname natexlab\endcsname\relax\def\natexlab#1{#1}\fi
	\expandafter\ifx\csname bibnamefont\endcsname\relax
	\def\bibnamefont#1{#1}\fi
	\expandafter\ifx\csname bibfnamefont\endcsname\relax
	\def\bibfnamefont#1{#1}\fi
	\expandafter\ifx\csname citenamefont\endcsname\relax
	\def\citenamefont#1{#1}\fi
	\expandafter\ifx\csname url\endcsname\relax
	\def\url#1{\texttt{#1}}\fi
	\expandafter\ifx\csname urlprefix\endcsname\relax\def\urlprefix{URL }\fi
	\providecommand{\bibinfo}[2]{#2}
	\providecommand{\eprint}[2][]{\url{#2}}
	
	\bibitem[{\citenamefont{Zeller and Pohl}(1971)}]{zellerpohl}
	\bibinfo{author}{\bibfnamefont{R.~C.} \bibnamefont{Zeller}} \bibnamefont{and}
	\bibinfo{author}{\bibfnamefont{R.~O.} \bibnamefont{Pohl}},
	\bibinfo{journal}{Phys. Rev. B} \textbf{\bibinfo{volume}{4}},
	\bibinfo{pages}{2029} (\bibinfo{year}{1971}).
	
	\bibitem[{\citenamefont{Yu and Leggett}(1988)}]{1988-Y.Leggett-CommCondMatPhys}
	\bibinfo{author}{\bibfnamefont{C.}~\bibnamefont{Yu}} \bibnamefont{and}
	\bibinfo{author}{\bibfnamefont{A.}~\bibnamefont{Leggett}},
	\bibinfo{journal}{Comments on Conden. Matter Phys.}
	\textbf{\bibinfo{volume}{14}}, \bibinfo{pages}{231} (\bibinfo{year}{1988}).
	
	\bibitem[{\citenamefont{Sokolov et~al.}(1997)\citenamefont{Sokolov, Calemczuk,
			Salce, Kisliuk, Quitmann, and Duval}}]{1997-AP.Sokolov-PRL}
	\bibinfo{author}{\bibfnamefont{A.~P.} \bibnamefont{Sokolov}},
	\bibinfo{author}{\bibfnamefont{R.}~\bibnamefont{Calemczuk}},
	\bibinfo{author}{\bibfnamefont{B.}~\bibnamefont{Salce}},
	\bibinfo{author}{\bibfnamefont{A.}~\bibnamefont{Kisliuk}},
	\bibinfo{author}{\bibfnamefont{D.}~\bibnamefont{Quitmann}}, \bibnamefont{and}
	\bibinfo{author}{\bibfnamefont{E.}~\bibnamefont{Duval}},
	\bibinfo{journal}{Phys. Rev. Lett.} \textbf{\bibinfo{volume}{78}},
	\bibinfo{pages}{2405} (\bibinfo{year}{1997}).
	
	\bibitem[{\citenamefont{Freeman and
			Anderson}(1986)}]{1986-JJ.Freeman-AC.Anderson-PRB}
	\bibinfo{author}{\bibfnamefont{J.~J.} \bibnamefont{Freeman}} \bibnamefont{and}
	\bibinfo{author}{\bibfnamefont{A.~C.} \bibnamefont{Anderson}},
	\bibinfo{journal}{Phys.Rev.B} \textbf{\bibinfo{volume}{34}},
	\bibinfo{pages}{5684} (\bibinfo{year}{1986}).
	
	\bibitem[{\citenamefont{Schirmacher}(2006)}]{2006-W.Schirmacher-EL}
	\bibinfo{author}{\bibfnamefont{W.}~\bibnamefont{Schirmacher}},
	\bibinfo{journal}{Europhys. Lett.} \textbf{\bibinfo{volume}{73}},
	\bibinfo{pages}{892} (\bibinfo{year}{2006}).
	
	\bibitem[{\citenamefont{Buchenau et~al.}(1986)\citenamefont{Buchenau, Prager,
			N\"ucker, Dianoux, Ahmad, and Phillips}}]{Buchenau-PRB-1986}
	\bibinfo{author}{\bibfnamefont{U.}~\bibnamefont{Buchenau}},
	\bibinfo{author}{\bibfnamefont{M.}~\bibnamefont{Prager}},
	\bibinfo{author}{\bibfnamefont{N.}~\bibnamefont{N\"ucker}},
	\bibinfo{author}{\bibfnamefont{A.~J.} \bibnamefont{Dianoux}},
	\bibinfo{author}{\bibfnamefont{N.}~\bibnamefont{Ahmad}}, \bibnamefont{and}
	\bibinfo{author}{\bibfnamefont{W.~A.} \bibnamefont{Phillips}},
	\bibinfo{journal}{Phys. Rev. B} \textbf{\bibinfo{volume}{34}},
	\bibinfo{pages}{5665} (\bibinfo{year}{1986}).
	
	\bibitem[{\citenamefont{Malinovsky and
			Sokolov}(1986)}]{1986-AP.Sokolov-SolidStateCom}
	\bibinfo{author}{\bibfnamefont{V.~K.} \bibnamefont{Malinovsky}}
	\bibnamefont{and} \bibinfo{author}{\bibfnamefont{A.~P.}
		\bibnamefont{Sokolov}}, \bibinfo{journal}{Solid State Commun.}
	\textbf{\bibinfo{volume}{57}}, \bibinfo{pages}{757} (\bibinfo{year}{1986}).
	
	\bibitem[{\citenamefont{Schirmacher et~al.}(1998)\citenamefont{Schirmacher,
			Diezemann, and Ganter}}]{1998-W.Schirmacher-PRL}
	\bibinfo{author}{\bibfnamefont{W.}~\bibnamefont{Schirmacher}},
	\bibinfo{author}{\bibfnamefont{G.}~\bibnamefont{Diezemann}},
	\bibnamefont{and} \bibinfo{author}{\bibfnamefont{C.}~\bibnamefont{Ganter}},
	\bibinfo{journal}{Phys. Rev. Lett.} \textbf{\bibinfo{volume}{81}},
	\bibinfo{pages}{136} (\bibinfo{year}{1998}).
	
	\bibitem[{\citenamefont{Shintani and Tanaka}(2008)}]{2008-H.Tanaka-NatureMat}
	\bibinfo{author}{\bibfnamefont{H.}~\bibnamefont{Shintani}} \bibnamefont{and}
	\bibinfo{author}{\bibfnamefont{H.}~\bibnamefont{Tanaka}},
	\bibinfo{journal}{Nat. Mater.} \textbf{\bibinfo{volume}{7}},
	\bibinfo{pages}{870} (\bibinfo{year}{2008}).
	
	\bibitem[{\citenamefont{Monaco and Mossa}(2009)}]{2009-G.Monaco-PNAS}
	\bibinfo{author}{\bibfnamefont{G.}~\bibnamefont{Monaco}} \bibnamefont{and}
	\bibinfo{author}{\bibfnamefont{S.}~\bibnamefont{Mossa}},
	\bibinfo{journal}{PNAS} \textbf{\bibinfo{volume}{106}},
	\bibinfo{pages}{16907} (\bibinfo{year}{2009}).
	
	\bibitem[{\citenamefont{Marruzzo et~al.}(2013)\citenamefont{Marruzzo,
			Schirmacher, Fratalocchi, and Ruocco}}]{2013-W.Schirmacher-ScientificRep}
	\bibinfo{author}{\bibfnamefont{A.}~\bibnamefont{Marruzzo}},
	\bibinfo{author}{\bibfnamefont{W.}~\bibnamefont{Schirmacher}},
	\bibinfo{author}{\bibfnamefont{A.}~\bibnamefont{Fratalocchi}},
	\bibnamefont{and} \bibinfo{author}{\bibfnamefont{G.}~\bibnamefont{Ruocco}},
	\bibinfo{journal}{Sci. Rep.} \textbf{\bibinfo{volume}{3}},
	\bibinfo{pages}{1142} (\bibinfo{year}{2013}).
	
	\bibitem[{\citenamefont{J\"ackle}(1981)}]{Jackle1981}
	\bibinfo{author}{\bibfnamefont{J.}~\bibnamefont{J\"ackle}}, in
	\emph{\bibinfo{booktitle}{Amorphous Solids: Low-Temperature Properties}},
	edited by \bibinfo{editor}{\bibfnamefont{W.~A.} \bibnamefont{Phillips}}
	(\bibinfo{publisher}{Springer, Berlin}, \bibinfo{year}{1981}), p.
	\bibinfo{pages}{135}.
	
	\bibitem[{\citenamefont{Buchenau et~al.}(1984)\citenamefont{Buchenau, N\"ucker,
			and Dianoux}}]{Buchenau-PRL-1984}
	\bibinfo{author}{\bibfnamefont{U.}~\bibnamefont{Buchenau}},
	\bibinfo{author}{\bibfnamefont{N.}~\bibnamefont{N\"ucker}}, \bibnamefont{and}
	\bibinfo{author}{\bibfnamefont{A.~J.} \bibnamefont{Dianoux}},
	\bibinfo{journal}{Phys. Rev. Lett.} \textbf{\bibinfo{volume}{53}},
	\bibinfo{pages}{2316} (\bibinfo{year}{1984}).
	
	\bibitem[{\citenamefont{Laird and Schober}(1991)}]{1991-Laird-PRL}
	\bibinfo{author}{\bibfnamefont{B.~B.} \bibnamefont{Laird}} \bibnamefont{and}
	\bibinfo{author}{\bibfnamefont{H.~R.} \bibnamefont{Schober}},
	\bibinfo{journal}{Phys. Rev. Lett.} \textbf{\bibinfo{volume}{66}},
	\bibinfo{pages}{636} (\bibinfo{year}{1991}).
	
	\bibitem[{\citenamefont{Baggioli and Zaccone}(2019)}]{Baggioli2019universal}
	\bibinfo{author}{\bibfnamefont{M.}~\bibnamefont{Baggioli}} \bibnamefont{and}
	\bibinfo{author}{\bibfnamefont{A.}~\bibnamefont{Zaccone}},
	\bibinfo{journal}{Phys. Rev. Lett.} \textbf{\bibinfo{volume}{122}},
	\bibinfo{pages}{145501} (\bibinfo{year}{2019}).
	
	\bibitem[{\citenamefont{Baggioli et~al.}(2019)\citenamefont{Baggioli, Milkus,
			and Zaccone}}]{Baggioli2019vibrational}
	\bibinfo{author}{\bibfnamefont{M.}~\bibnamefont{Baggioli}},
	\bibinfo{author}{\bibfnamefont{R.}~\bibnamefont{Milkus}}, \bibnamefont{and}
	\bibinfo{author}{\bibfnamefont{A.}~\bibnamefont{Zaccone}},
	\bibinfo{journal}{Phys. Rev. E} \textbf{\bibinfo{volume}{100}},
	\bibinfo{pages}{062131} (\bibinfo{year}{2019}).
	
	\bibitem[{\citenamefont{Karpov et~al.}(1983)\citenamefont{Karpov, Klinger, and
			Ignatiev}}]{karpov83}
	\bibinfo{author}{\bibfnamefont{V.~G.} \bibnamefont{Karpov}},
	\bibinfo{author}{\bibfnamefont{M.~I.} \bibnamefont{Klinger}},
	\bibnamefont{and} \bibinfo{author}{\bibfnamefont{F.~N.}
		\bibnamefont{Ignatiev}}, \bibinfo{journal}{Sov. Phys. JETP}
	\textbf{\bibinfo{volume}{57}}, \bibinfo{pages}{439} (\bibinfo{year}{1983}).
	
	\bibitem[{\citenamefont{Buchenau et~al.}(1991)\citenamefont{Buchenau, Galperin,
			Gurevich, and Schober}}]{Buchenau91a}
	\bibinfo{author}{\bibfnamefont{U.}~\bibnamefont{Buchenau}},
	\bibinfo{author}{\bibfnamefont{Y.~M.} \bibnamefont{Galperin}},
	\bibinfo{author}{\bibfnamefont{V.~L.} \bibnamefont{Gurevich}},
	\bibnamefont{and} \bibinfo{author}{\bibfnamefont{H.~R.}
		\bibnamefont{Schober}}, \bibinfo{journal}{Phys. Rev. B}
	\textbf{\bibinfo{volume}{43}}, \bibinfo{pages}{5039} (\bibinfo{year}{1991}).
	
	\bibitem[{\citenamefont{Gurevich et~al.}(2003)\citenamefont{Gurevich, Parshin,
			and Schober}}]{gurevich03a}
	\bibinfo{author}{\bibfnamefont{V.~L.} \bibnamefont{Gurevich}},
	\bibinfo{author}{\bibfnamefont{D.~A.} \bibnamefont{Parshin}},
	\bibnamefont{and} \bibinfo{author}{\bibfnamefont{H.~R.}
		\bibnamefont{Schober}}, \bibinfo{journal}{Phys. Rev. B}
	\textbf{\bibinfo{volume}{67}} (\bibinfo{year}{2003}).
	
	\bibitem[{\citenamefont{Mart\'in-Mayor
			et~al.}(2000)\citenamefont{Mart\'in-Mayor, Parisi, and
			Verocchio}}]{martin-mayor00}
	\bibinfo{author}{\bibfnamefont{V.}~\bibnamefont{Mart\'in-Mayor}},
	\bibinfo{author}{\bibfnamefont{G.}~\bibnamefont{Parisi}}, \bibnamefont{and}
	\bibinfo{author}{\bibfnamefont{P.}~\bibnamefont{Verocchio}},
	\bibinfo{journal}{Phys. Rev. E} \textbf{\bibinfo{volume}{62}},
	\bibinfo{pages}{2373} (\bibinfo{year}{2000}).
	
	\bibitem[{\citenamefont{Taraskin et~al.}(2001)\citenamefont{Taraskin, Loh,
			Natarajan, and Elliott}}]{2001-SN.Taraskin-PRL}
	\bibinfo{author}{\bibfnamefont{S.~N.} \bibnamefont{Taraskin}},
	\bibinfo{author}{\bibfnamefont{Y.~L.} \bibnamefont{Loh}},
	\bibinfo{author}{\bibfnamefont{G.}~\bibnamefont{Natarajan}},
	\bibnamefont{and} \bibinfo{author}{\bibfnamefont{S.~R.}
		\bibnamefont{Elliott}}, \bibinfo{journal}{Phys. Rev. Lett.}
	\textbf{\bibinfo{volume}{86}}, \bibinfo{pages}{1255} (\bibinfo{year}{2001}).
	
	\bibitem[{\citenamefont{Schirmacher et~al.}(2007)\citenamefont{Schirmacher,
			Ruocco, and Scopigno}}]{2007-W.Schirmacher-PRL}
	\bibinfo{author}{\bibfnamefont{W.}~\bibnamefont{Schirmacher}},
	\bibinfo{author}{\bibfnamefont{G.}~\bibnamefont{Ruocco}}, \bibnamefont{and}
	\bibinfo{author}{\bibfnamefont{T.}~\bibnamefont{Scopigno}},
	\bibinfo{journal}{Phys. Rev. Lett.} \textbf{\bibinfo{volume}{98}},
	\bibinfo{pages}{95} (\bibinfo{year}{2007}).
	
	\bibitem[{\citenamefont{Schirmacher}(2013)}]{2013-W.Schirmacher-PSSB}
	\bibinfo{author}{\bibfnamefont{W.}~\bibnamefont{Schirmacher}},
	\bibinfo{journal}{Phys. Stat. Sol. (b)} \textbf{\bibinfo{volume}{250}},
	\bibinfo{pages}{937} (\bibinfo{year}{2013}).
	
	\bibitem[{\citenamefont{Schirmacher}(2014)}]{schi14}
	\bibinfo{author}{\bibfnamefont{W.}~\bibnamefont{Schirmacher}},
	\bibinfo{journal}{J. Noncryst. Sol.} \textbf{\bibinfo{volume}{407}},
	\bibinfo{pages}{133} (\bibinfo{year}{2014}).
	
	\bibitem[{\citenamefont{Chumakov et~al.}(2011)\citenamefont{Chumakov, Monaco,
			Monaco, Crichton, Bosak, R\"{u}ffer, Meyer, Kargl, Comez, and
			Fioretto}}]{2011-Chumakov-PRL}
	\bibinfo{author}{\bibfnamefont{A.~I.} \bibnamefont{Chumakov}},
	\bibinfo{author}{\bibfnamefont{G.}~\bibnamefont{Monaco}},
	\bibinfo{author}{\bibfnamefont{A.}~\bibnamefont{Monaco}},
	\bibinfo{author}{\bibfnamefont{W.~A.} \bibnamefont{Crichton}},
	\bibinfo{author}{\bibfnamefont{A.}~\bibnamefont{Bosak}},
	\bibinfo{author}{\bibfnamefont{R.}~\bibnamefont{R\"{u}ffer}},
	\bibinfo{author}{\bibfnamefont{A.}~\bibnamefont{Meyer}},
	\bibinfo{author}{\bibfnamefont{F.}~\bibnamefont{Kargl}},
	\bibinfo{author}{\bibfnamefont{L.}~\bibnamefont{Comez}}, \bibnamefont{and}
	\bibinfo{author}{\bibfnamefont{D.}~\bibnamefont{Fioretto}},
	\bibinfo{journal}{Phys. Rev. Lett.} \textbf{\bibinfo{volume}{106}},
	\bibinfo{pages}{1109} (\bibinfo{year}{2011}).
	
	\bibitem[{\citenamefont{Chumakov et~al.}(2014)\citenamefont{Chumakov, Monaco,
			Fontana, Bosak, Hermann, Bessas, Wehinger, Crichton, Krisch, and
			R\"{u}ffer}}]{2014-Chumakov-PRL}
	\bibinfo{author}{\bibfnamefont{A.~I.} \bibnamefont{Chumakov}},
	\bibinfo{author}{\bibfnamefont{G.}~\bibnamefont{Monaco}},
	\bibinfo{author}{\bibfnamefont{A.}~\bibnamefont{Fontana}},
	\bibinfo{author}{\bibfnamefont{A.}~\bibnamefont{Bosak}},
	\bibinfo{author}{\bibfnamefont{R.~P.} \bibnamefont{Hermann}},
	\bibinfo{author}{\bibfnamefont{D.}~\bibnamefont{Bessas}},
	\bibinfo{author}{\bibfnamefont{B.}~\bibnamefont{Wehinger}},
	\bibinfo{author}{\bibfnamefont{W.~A.} \bibnamefont{Crichton}},
	\bibinfo{author}{\bibfnamefont{M.}~\bibnamefont{Krisch}}, \bibnamefont{and}
	\bibinfo{author}{\bibfnamefont{R.}~\bibnamefont{R\"{u}ffer}},
	\bibinfo{journal}{Phys. Rev. Lett.} \textbf{\bibinfo{volume}{112}},
	\bibinfo{pages}{339} (\bibinfo{year}{2014}).
	
	\bibitem[{\citenamefont{Zorn}(2011)}]{zorn11}
	\bibinfo{author}{\bibfnamefont{R.}~\bibnamefont{Zorn}},
	\bibinfo{journal}{Physics} \textbf{\bibinfo{volume}{4}}, \bibinfo{pages}{44}
	(\bibinfo{year}{2011}).
	
	\bibitem[{\citenamefont{Wang et~al.}(2018)\citenamefont{Wang, Hong, Wang,
			Schirmacher, and Zhang}}]{Wang2018-disentangling}
	\bibinfo{author}{\bibfnamefont{Y.}~\bibnamefont{Wang}},
	\bibinfo{author}{\bibfnamefont{L.}~\bibnamefont{Hong}},
	\bibinfo{author}{\bibfnamefont{Y.}~\bibnamefont{Wang}},
	\bibinfo{author}{\bibfnamefont{W.}~\bibnamefont{Schirmacher}},
	\bibnamefont{and} \bibinfo{author}{\bibfnamefont{J.}~\bibnamefont{Zhang}},
	\bibinfo{journal}{Phys. Rev. B} \textbf{\bibinfo{volume}{98}},
	\bibinfo{pages}{174207} (\bibinfo{year}{2018}).
	
	\bibitem[{\citenamefont{Binder and Kob}(2011)}]{2011-K.Binder-W.Kob-GlassBook}
	\bibinfo{author}{\bibfnamefont{K.}~\bibnamefont{Binder}} \bibnamefont{and}
	\bibinfo{author}{\bibfnamefont{W.}~\bibnamefont{Kob}},
	\emph{\bibinfo{title}{Glassy materials and disordered solids}}
	(\bibinfo{publisher}{World Scientific,}, \bibinfo{year}{2011}).
	
	\bibitem[{\citenamefont{Chen et~al.}(2013)\citenamefont{Chen, Still,
			Schoenholz, Aptowicz, Schindler, Maggs, Liu, and Yodh}}]{2013-K.Chen-PRE}
	\bibinfo{author}{\bibfnamefont{K.}~\bibnamefont{Chen}},
	\bibinfo{author}{\bibfnamefont{T.}~\bibnamefont{Still}},
	\bibinfo{author}{\bibfnamefont{S.}~\bibnamefont{Schoenholz}},
	\bibinfo{author}{\bibfnamefont{K.~B.} \bibnamefont{Aptowicz}},
	\bibinfo{author}{\bibfnamefont{M.}~\bibnamefont{Schindler}},
	\bibinfo{author}{\bibfnamefont{A.~C.} \bibnamefont{Maggs}},
	\bibinfo{author}{\bibfnamefont{A.~J.} \bibnamefont{Liu}}, \bibnamefont{and}
	\bibinfo{author}{\bibfnamefont{A.~G.} \bibnamefont{Yodh}},
	\bibinfo{journal}{Phys. Rev. E} \textbf{\bibinfo{volume}{88}},
	\bibinfo{pages}{3388} (\bibinfo{year}{2013}).
	
	\bibitem[{\citenamefont{Tanaka et~al.}(2010)\citenamefont{Tanaka, Kawasaki,
			Shintani, and Watanabe}}]{2010-H.Tanaka-NatrueMat}
	\bibinfo{author}{\bibfnamefont{H.}~\bibnamefont{Tanaka}},
	\bibinfo{author}{\bibfnamefont{T.}~\bibnamefont{Kawasaki}},
	\bibinfo{author}{\bibfnamefont{H.}~\bibnamefont{Shintani}}, \bibnamefont{and}
	\bibinfo{author}{\bibfnamefont{K.}~\bibnamefont{Watanabe}},
	\bibinfo{journal}{Nat. Mater.} \textbf{\bibinfo{volume}{9}},
	\bibinfo{pages}{324} (\bibinfo{year}{2010}).
	
	\bibitem[{\citenamefont{Bi et~al.}(2011)\citenamefont{Bi, Zhang, Chakraborty,
			and Behringer}}]{2011-JZ-Nature}
	\bibinfo{author}{\bibfnamefont{D.}~\bibnamefont{Bi}},
	\bibinfo{author}{\bibfnamefont{J.}~\bibnamefont{Zhang}},
	\bibinfo{author}{\bibfnamefont{B.}~\bibnamefont{Chakraborty}},
	\bibnamefont{and} \bibinfo{author}{\bibfnamefont{R.~P.}
		\bibnamefont{Behringer}}, \bibinfo{journal}{Nature}
	\textbf{\bibinfo{volume}{480}}, \bibinfo{pages}{355} (\bibinfo{year}{2011}).
	
	\bibitem[{\citenamefont{Zhang et~al.}(2017)\citenamefont{Zhang, Zheng, Wang,
			Zhang, Jin, Hong, Wang, and Zhang}}]{2017-LZhang-NatCommun}
	\bibinfo{author}{\bibfnamefont{L.}~\bibnamefont{Zhang}},
	\bibinfo{author}{\bibfnamefont{J.}~\bibnamefont{Zheng}},
	\bibinfo{author}{\bibfnamefont{Y.}~\bibnamefont{Wang}},
	\bibinfo{author}{\bibfnamefont{L.}~\bibnamefont{Zhang}},
	\bibinfo{author}{\bibfnamefont{Z.}~\bibnamefont{Jin}},
	\bibinfo{author}{\bibfnamefont{L.}~\bibnamefont{Hong}},
	\bibinfo{author}{\bibfnamefont{Y.}~\bibnamefont{Wang}}, \bibnamefont{and}
	\bibinfo{author}{\bibfnamefont{J.}~\bibnamefont{Zhang}},
	\bibinfo{journal}{Nat. Commun.} \textbf{\bibinfo{volume}{8}},
	\bibinfo{pages}{67} (\bibinfo{year}{2017}).
	
	\bibitem[{\citenamefont{Gao et~al.}(2006)\citenamefont{Gao, B{\l}awzdziewicz,
			and O'Hern}}]{2006-Ohern-PRE}
	\bibinfo{author}{\bibfnamefont{G.~J.} \bibnamefont{Gao}},
	\bibinfo{author}{\bibfnamefont{J.}~\bibnamefont{B{\l}awzdziewicz}},
	\bibnamefont{and} \bibinfo{author}{\bibfnamefont{C.~S.}
		\bibnamefont{O'Hern}}, \bibinfo{journal}{Phys. Rev. E}
	\textbf{\bibinfo{volume}{74}}, \bibinfo{pages}{061304}
	(\bibinfo{year}{2006}).
	
	\bibitem[{\citenamefont{Silbert et~al.}(2005)\citenamefont{Silbert, Liu, and
			Nagel}}]{2005-LE.Silbert-PRL}
	\bibinfo{author}{\bibfnamefont{L.~E.} \bibnamefont{Silbert}},
	\bibinfo{author}{\bibfnamefont{A.~J.} \bibnamefont{Liu}}, \bibnamefont{and}
	\bibinfo{author}{\bibfnamefont{S.~R.} \bibnamefont{Nagel}},
	\bibinfo{journal}{Phys. Rev. Lett} \textbf{\bibinfo{volume}{95}}
	(\bibinfo{year}{2005}).
	
	\bibitem[{\citenamefont{Anderson}(1958)}]{1958-PW.Anderson-PhyscalReview}
	\bibinfo{author}{\bibfnamefont{P.~W.} \bibnamefont{Anderson}},
	\bibinfo{journal}{Phys. Rev.} \textbf{\bibinfo{volume}{109}},
	\bibinfo{pages}{1492} (\bibinfo{year}{1958}).
	
	\bibitem[{\citenamefont{Hu et~al.}(2008)\citenamefont{Hu, Strybulevych, Page,
			Skipetrov, and van Tiggelen}}]{2008-HefeiHu-NatPhys}
	\bibinfo{author}{\bibfnamefont{H.}~\bibnamefont{Hu}},
	\bibinfo{author}{\bibfnamefont{A.}~\bibnamefont{Strybulevych}},
	\bibinfo{author}{\bibfnamefont{J.~H.} \bibnamefont{Page}},
	\bibinfo{author}{\bibfnamefont{S.~E.} \bibnamefont{Skipetrov}},
	\bibnamefont{and} \bibinfo{author}{\bibfnamefont{B.~A.} \bibnamefont{van
			Tiggelen}}, \bibinfo{journal}{Nat. Phys.} \textbf{\bibinfo{volume}{4}},
	\bibinfo{pages}{945} (\bibinfo{year}{2008}).
	
	\bibitem[{\citenamefont{Huang and Wu}(2009)}]{2009-BJ.Huang-TM.Wu-PRE}
	\bibinfo{author}{\bibfnamefont{B.~J.} \bibnamefont{Huang}} \bibnamefont{and}
	\bibinfo{author}{\bibfnamefont{T.~M.} \bibnamefont{Wu}},
	\bibinfo{journal}{Phys. Rev. E} \textbf{\bibinfo{volume}{79}},
	\bibinfo{pages}{041105} (\bibinfo{year}{2009}).
	
	\bibitem[{\citenamefont{Zhang et~al.}(2021)\citenamefont{Zhang, Wang, Zheng,
			Sun, and Zhang}}]{2021-LZhang-Level}
	\bibinfo{author}{\bibfnamefont{L.}~\bibnamefont{Zhang}},
	\bibinfo{author}{\bibfnamefont{Y.}~\bibnamefont{Wang}},
	\bibinfo{author}{\bibfnamefont{J.}~\bibnamefont{Zheng}},
	\bibinfo{author}{\bibfnamefont{A.}~\bibnamefont{Sun}}, \bibnamefont{and}
	\bibinfo{author}{\bibfnamefont{J.}~\bibnamefont{Zhang}},
	\bibinfo{journal}{Phys. Rev. B} \textbf{\bibinfo{volume}{103}},
	\bibinfo{pages}{104201} (\bibinfo{year}{2021}).
	
	\bibitem[{\citenamefont{Pinski et~al.}(2012{\natexlab{a}})\citenamefont{Pinski,
			Schirmacher, Whall, and R\"{o}mer}}]{2012-W.Schirmacher-J.PhysCondensMatter}
	\bibinfo{author}{\bibfnamefont{S.~D.} \bibnamefont{Pinski}},
	\bibinfo{author}{\bibfnamefont{W.}~\bibnamefont{Schirmacher}},
	\bibinfo{author}{\bibfnamefont{T.}~\bibnamefont{Whall}}, \bibnamefont{and}
	\bibinfo{author}{\bibfnamefont{R.~A.} \bibnamefont{R\"{o}mer}},
	\bibinfo{journal}{J. Phys.: Condens. Matter} \textbf{\bibinfo{volume}{24}},
	\bibinfo{pages}{4172} (\bibinfo{year}{2012}{\natexlab{a}}).
	
	\bibitem[{\citenamefont{Pinski et~al.}(2012{\natexlab{b}})\citenamefont{Pinski,
			Schirmacher, and R\"{o}mer}}]{2012-W.Schirmacher-EPL}
	\bibinfo{author}{\bibfnamefont{S.~D.} \bibnamefont{Pinski}},
	\bibinfo{author}{\bibfnamefont{W.}~\bibnamefont{Schirmacher}},
	\bibnamefont{and} \bibinfo{author}{\bibfnamefont{R.~A.}
		\bibnamefont{R\"{o}mer}}, \bibinfo{journal}{Europhys. Lett.}
	\textbf{\bibinfo{volume}{97}}, \bibinfo{pages}{16007}
	(\bibinfo{year}{2012}{\natexlab{b}}).
	
	\bibitem[{\citenamefont{Tong et~al.}(2015)\citenamefont{Tong, Tan, and
			Xu}}]{2015-HTong-NXu-SciRep}
	\bibinfo{author}{\bibfnamefont{H.}~\bibnamefont{Tong}},
	\bibinfo{author}{\bibfnamefont{P.}~\bibnamefont{Tan}}, \bibnamefont{and}
	\bibinfo{author}{\bibfnamefont{N.}~\bibnamefont{Xu}}, \bibinfo{journal}{Sci.
		Rep.} \textbf{\bibinfo{volume}{5}} (\bibinfo{year}{2015}).
	
	\bibitem[{\citenamefont{Nie et~al.}(2017)\citenamefont{Nie, Tong, Liu, Zu, and
			Xu}}]{2017-YHNie-FrontPhys}
	\bibinfo{author}{\bibfnamefont{Y.}~\bibnamefont{Nie}},
	\bibinfo{author}{\bibfnamefont{H.}~\bibnamefont{Tong}},
	\bibinfo{author}{\bibfnamefont{J.}~\bibnamefont{Liu}},
	\bibinfo{author}{\bibfnamefont{M.}~\bibnamefont{Zu}}, \bibnamefont{and}
	\bibinfo{author}{\bibfnamefont{N.}~\bibnamefont{Xu}},
	\bibinfo{journal}{Front. Phys.} \textbf{\bibinfo{volume}{12}},
	\bibinfo{pages}{126301} (\bibinfo{year}{2017}).
	
	\bibitem[{\citenamefont{Tighe et~al.}(2008)\citenamefont{Tighe, van Eerd, and
			Vlugt}}]{2008-BP.tighe-PRL}
	\bibinfo{author}{\bibfnamefont{B.~P.} \bibnamefont{Tighe}},
	\bibinfo{author}{\bibfnamefont{A.~R.} \bibnamefont{van Eerd}},
	\bibnamefont{and} \bibinfo{author}{\bibfnamefont{T.~J.} \bibnamefont{Vlugt}},
	\bibinfo{journal}{Phys. Rev. Lett.} \textbf{\bibinfo{volume}{100}},
	\bibinfo{pages}{1151} (\bibinfo{year}{2008}).
	
	\bibitem[{\citenamefont{Schirmacher et~al.}(2008)\citenamefont{Schirmacher,
			Schmid, Tomaras, Viliani, Baldi, Ruocco, and
			Scopigno}}]{2008-W.Schirmacher-PSSB}
	\bibinfo{author}{\bibfnamefont{W.}~\bibnamefont{Schirmacher}},
	\bibinfo{author}{\bibfnamefont{B.}~\bibnamefont{Schmid}},
	\bibinfo{author}{\bibfnamefont{C.}~\bibnamefont{Tomaras}},
	\bibinfo{author}{\bibfnamefont{G.}~\bibnamefont{Viliani}},
	\bibinfo{author}{\bibfnamefont{G.}~\bibnamefont{Baldi}},
	\bibinfo{author}{\bibfnamefont{G.}~\bibnamefont{Ruocco}}, \bibnamefont{and}
	\bibinfo{author}{\bibfnamefont{T.}~\bibnamefont{Scopigno}},
	\bibinfo{journal}{Phys. Stat. Sol. (c)} \textbf{\bibinfo{volume}{5}},
	\bibinfo{pages}{862} (\bibinfo{year}{2008}).
	
\end{thebibliography}

\end{document}